\documentclass[12pt,a4paper]{article}

\usepackage{amssymb}
\usepackage{amsfonts}
\usepackage{amsthm}
\usepackage{mathrsfs}

\makeatletter
\newcommand{\lyxaddress}[1]{
\par {\raggedright #1
\vspace{1.4em}
\noindent\par}
}
\makeatother

\newcommand{\sH}{\mathscr{H}}
\newcommand{\sL}{\mathscr{L}}
\newcommand{\sI}{\mathscr{I}}
\newcommand{\sF}{\mathscr{F}}
\newcommand{\sD}{\mathscr{D}}
\newcommand{\sB}{\mathscr{B}}
\newcommand{\R}{\mathbb{R}}
\newcommand{\Z}{\mathbb{Z}}
\newcommand{\KK}{\mathcal{K}}
\renewcommand{\AA}{\mathcal{A}}
\newcommand{\BB}{\mathcal{B}}

\theoremstyle{remark}

\newtheorem*{rem*}{Remark}
\newtheorem*{rems*}{Remarks}

\theoremstyle{definition}

\newtheorem*{def*}{Definition}

\newcommand{\Ran}{\mathop\mathrm{Ran}\nolimits}
\newcommand{\Tr}{\mathop\mathrm{Tr}\nolimits}
\newcommand{\rot}{\mathop\mathrm{rot}\nolimits}
\newcommand{\dist}{\mathop\mathrm{dist}\nolimits}
\newcommand{\dd}{\mathrm{d}}

\begin{document}

\title{Propagators associated to periodic Hamiltonians: an example of
  the Aharonov-Bohm Hamiltonian with two vortices}

\author{P.~Koc\'abov\'a, P.~\v{S}\v{t}ov\'\i\v{c}ek}
\date{}

\maketitle

\lyxaddress{Department of Mathematics, Faculty of Nuclear Science,
  Czech Technical University, Trojanova 13, 120 00 Prague, Czech
  Republic}

\begin{abstract}\noindent
  We consider an invariant quantum Hamiltonian $H=-\Delta_{LB}+V$ in
  the $L^{2}$ space based on a Riemannian manifold $\tilde{M}$ with a
  discrete symmetry group $\Gamma$. Typically, $\tilde{M}$ is the
  universal covering space of a multiply connected manifold $M$ and
  $\Gamma$ is the fundamental group of $M$. To any unitary
  representation $\Lambda$ of $\Gamma$ one can relate another operator
  on $M=\tilde{M}/\Gamma$, called $H_\Lambda$, which formally
  corresponds to the same differential operator as $H$ but which is
  determined by quasi-periodic boundary conditions. We give a brief
  review of the Bloch decomposition of $H$ and of a formula relating
  the propagators associated to the Hamiltonians $H_\Lambda$ and $H$.
  Then we concentrate on the example of the Aharonov-Bohm effect with
  two vortices. We explain in detail the construction of the
  propagator in this case and indicate all essential intermediate
  steps.
  \vskip \baselineskip\noindent PACS numbers: 03.65.Db
\end{abstract}

\section{Introduction}

Suppose that there is given a connected Riemannian manifold
$\tilde{M}$ with a discrete symmetry group $\Gamma$. Let us consider a
$\Gamma$-periodic Hamilton operator in $L^2(\tilde{M})$ of the form
$H=-\Delta_{LB}+V$ where $\Delta_{LB}$ is the Laplace-Beltrami
operator and $V$ is a $\Gamma$-invariant bounded real function on
$\tilde{M}$. To any unitary representation $\Lambda$ of $\Gamma$ one
can relate another operator on $M=\tilde{M}/\Gamma$, called
$H_\Lambda$, which formally corresponds to the same differential
operator as $H$ but which is determined by quasi-periodic boundary
conditions.  In the framework of the Feynman path integral there was
derived a remarkable formula relating the propagators
$\KK^\Lambda_t(x,x_0)$ and $\KK_t(x,x_0)$ associated respectively to
the Hamiltonians $H_\Lambda$ and $H$ \cite{schulman1,schulman2}. An
analogous formula is also known for heat kernels \cite{atiyah}. There
exists also an opposite point of view when one decomposes the operator
$H$ into a direct integral with components $H_\Lambda$ where $\Lambda$
runs over all irreducible unitary representations of $\Gamma$
\cite{sunada,aschetal,gruber1}. The evolution operator then decomposes
correspondingly. This type of decomposition is an essential step in
the Bloch analysis. Let us also note that an alternative approach to
the Bloch analysis, based on a more algebraic point of view, has been
proposed recently in \cite{gruber2}.

The both relations, the propagator formula on the one hand and the
generalized Bloch decomposition on the other hand, are in a sense
mutually inverse \cite{kocabovastovicek}. In the current paper we give
a brief review of basic results concerning this relationship. The main
purpose of this contribution is, however, a detailed discussion of an
application of the formula for propagators. We consider the example of
the Aharonov-Bohm effect with two vortices. In this case $\tilde{M}$
is identified with the universal covering space of the plane with two
excluded points and $\Gamma$ is the fundamental group of the same
manifold. This problem has been already treated by one of the authors
quite a long time ago in \cite{pla89}. But even in more recent papers
one can encounter a discussion of the problem itself
\cite{mashkevichetal} as well as of the involved methods
\cite{hannaythain,giraudetal}. It should be stressed that article
\cite{pla89} is a very brief letter in which only the final formula is
presented without any detailed hints about its derivation. This is so
even though some steps in the derivation are rather complicated and in
no way obvious. Since these details were never published anywhere and
since without them the resulting formula may look a bit obscure we aim
here to fill in this gap by explaining the approach more carefully and
by indicating the necessary intermediate steps.

The paper is organized as follows. In
Section~\ref{sec:propag_assoc_periodH} we give a brief review of basic
results concerning the relationship between the generalized Bloch
analysis and the formula for propagators associated to periodic
Hamiltonians. In Section~\ref{sec:ABpropag_cover} we explain the
construction of the propagator on the universal covering space in the
case of the Aharonov-Bohm effect with two vortices. In
Section~\ref{sec:ABpropag_Lambda} we discuss the application of the
propagator formula in this particular case.

\section{Propagators associated to periodic Hamiltonians}
\label{sec:propag_assoc_periodH}

\subsection{Periodic Hamiltonians}
\label{sec:periodic_H}

Let $\tilde{M}$ be a connected Riemannian manifold with a discrete and
at most countable symmetry group $\Gamma$. The action of $\Gamma$ on
$\tilde{M}$ is assumed to be smooth, free and proper (also called
properly discontinuous). Denote by $\tilde{\mu}$ the measure on
$\tilde{M}$ induced by the Riemannian metric. The quotient
$M=\tilde{M}/\Gamma$ is a connected Riemannian manifold with an
induced measure $\mu$. This way one gets a principal fiber bundle
$\pi:\tilde{M}\to{}M$ with the structure group $\Gamma$. The $L^2$
spaces on the manifolds $M$ and $\tilde{M}$ are everywhere tacitly
understood with the measures $\mu$ and $\tilde\mu$, respectively.

Typically, $\tilde{M}$ is the universal covering space of $M$ and
$\Gamma=\pi_{1}(M)$ is the fundamental group of $M$. For example, this
is the case when one is considering the Aharonov-Bohm effect.

To a unitary representation $\Lambda$ of $\Gamma$ in a separable
Hilbert space $\sL_{\Lambda}$ one relates the Hilbert space
$\sH_{\Lambda}$ formed by $\Lambda$-equivariant vector-valued
functions on $\tilde{M}$. This means that any function
$\psi\in\sH_{\Lambda}$ is measurable with values in $\sL_{\Lambda}$
and satisfies
\begin{displaymath}
  \forall s\in\Gamma,\,
  \psi(s\cdot y)=\Lambda(s)\psi(y)\textrm{~~almost everywhere on~}
  \tilde{M}.
\end{displaymath}
Moreover, the norm of $\psi$ induced by the scalar product is finite.
If $\psi_{1},\psi_{2}\in\sH_{\Lambda}$ then the function
$y\mapsto\langle\psi_{1}(y),\psi_{2}(y)\rangle$ defined on $\tilde{M}$
is $\Gamma$-invariant and so it projects to a function
$s_{\psi_{1},\psi_{2}}$ defined on $M$, and the scalar product is
defined by
\begin{displaymath}
  \langle\psi_{1},\psi_{2}\rangle
  = \int_{M}s_{\psi_{1},\psi_{2}}(x)\,d\mu(x).
\end{displaymath}

As already announced, our discussion concerns $\Gamma$-periodic
Hamiltonians on $\tilde{M}$ of the form $H=-\Delta_{LB}+V$ where
$\Delta_{LB}$ is the Laplace-Beltrami operator and $V(y)$ is a
$\Gamma$-invariant measurable bounded real function on $\tilde{M}$.
Here we accept the Friedrichs extension as the preferred self-adjoint
extension of semibounded symmetric operators defined on test
functions.

To the same differential operator, $-\Delta_{LB}+V$, one can relate a
selfadjoint operator $H_{\Lambda}$ in the space $\sH_{\Lambda}$ for
any unitary representation $\Lambda$ of $\Gamma$. Let us define
$\Phi_{\Lambda}:C_{0}^{\infty}(\tilde{M})\otimes\sL_{\Lambda}
\to\sH_{\Lambda}$
by
\begin{displaymath}
  \forall\varphi\in C_{0}^{\infty}(\tilde{M}),
  \forall v\in\sL_{\Lambda},\textrm{~}
  \left(\Phi_{\Lambda}\varphi\otimes v\right)(y)
  = \sum_{s\in\Gamma}\varphi(s\cdot y)\,\Lambda(s^{-1})v.
\end{displaymath}
Since the action of $\Gamma$ is proper, the vector-valued function
$\Phi_{\Lambda}\,\varphi\otimes{}v$ is smooth. Moreover,
$\Phi_{\Lambda}\,\varphi\otimes{}v$ is $\Lambda$-equivariant and the
norm of $\Phi_{\Lambda}\,\varphi\otimes{}v$ in $\sH_{\Lambda}$ is
finite. Furthermore, the range of $\Phi_{\Lambda}$ is dense in
$\sH_{\Lambda}$. The Laplace-Beltrami operator is well defined on
$\Ran(\Phi_{\Lambda})$ and it holds
\begin{displaymath}
  \Delta_{LB}\Phi_{\Lambda}[\varphi\otimes v]
  = \Phi_{\Lambda}[\Delta_{LB}\varphi\otimes v].
\end{displaymath}
One can also verify that the differential operator $-\Delta_{LB}$ is
positive on the domain $\Ran(\Phi_{\Lambda})\subset\sH_{\Lambda}$.
Since the function $V(y)$ is $\Gamma$-invariant, the multiplication
operator by $V$ is well defined in the Hilbert space $\sH_{\Lambda}$.
The Hamiltonian $H_{\Lambda}$ is defined as the Friedrichs extension
of the differential operator $-\Delta_{LB}+V$ considered on the domain
$\Ran\Phi_{\Lambda}$.

\subsection{A generalization of the Bloch analysis}
\label{sec:bloch}

Let $\hat{\Gamma}$ be the dual space to $\Gamma$ (the quotient space
of the space of irreducible unitary representations of $\Gamma$). In
the first step of the generalized Bloch analysis one decomposes $H$
into a direct integral over $\hat{\Gamma}$ with the components being
equal to $H_\Lambda$. As a corollary one obtains a similar
relationship for the evolution operators $U(t)=\exp(-itH)$ and
$U_{\Lambda}(t)=\exp(-itH_{\Lambda})$, $t\in\mathbb{R}$. To achieve
this goal a well defined harmonic analysis on the group $\Gamma$ is
necessary.

It is known that the harmonic analysis is well established for locally
compact groups of type~I \cite{shtern}. So all formulas presented
bellow are perfectly well defined provided $\Gamma$ is a type~I group.
A countable discrete group is type~I, however, if and only if it has
an Abelian normal subgroup of finite index \cite[Satz~6]{thoma}. This
means that there exist multiply connected configuration spaces of
interest whose fundamental groups are not of type~I. For example, the
fundamental group in the case of the Aharonov-Bohm effect with two
vortices is the free group with two generators and it is not of
type~I.  Fortunately, in this case, too, there exists a well defined
harmonic analysis \cite{talamanca}.

Let us recall the basic properties of the harmonic analysis on
discrete type~I groups \cite{shtern}. In that case the Haar measure on
$\Gamma$ is chosen as the counting measure. Let $\mathrm{d}\hat{m}$ be
the Plancherel measure on $\hat{\Gamma}$. Denote by
$\sI_{2}(\sL_{\Lambda})\equiv\sL_{\Lambda}^{\ast}\otimes\sL_{\Lambda}$
the Hilbert space formed by Hilbert-Schmidt operators on
$\sL_{\Lambda}$ ($\sL_{\Lambda}^{\ast}$ is the dual space to
$\sL_{\Lambda}$). The Fourier transformation is defined as a unitary
mapping
\[
\sF:L^{2}(\Gamma)\to\int_{\hat{\Gamma}}^{\oplus}
\sI_{2}(\sL_{\Lambda})\,\mathrm{d}\hat{m}(\Lambda).
\]
For $f\in L^{1}(\Gamma)\subset L^{2}(\Gamma)$ one has
\[
\sF[f](\Lambda)=\sum_{s\in\Gamma}f(s)\Lambda(s).
\]
Conversely, if $f$ is of the form $f=g\ast{}h$ (the convolution) where
$g,h\in L^{1}(\Gamma)$, and $\hat{f}=\sF[f]$ then
\[
f(s)=\int_{\hat{\Gamma}}\Tr[\Lambda(s)^{\ast}\hat{f}(\Lambda)]\,
\mathrm{d}\hat{m}(\Lambda).
\]

It is known that if $\Gamma$ is a countable discrete group of type~I
then $\dim\sL_{\Lambda}$ is a bounded function of $\Lambda$ on the
dual space $\hat{\Gamma}$ \cite[Korollar~I]{thoma}. Using the
unitarity of the Fourier transform one finds that
\begin{displaymath}
  \hat{m}(\hat{\Gamma}) \leq
  \int_{\hat{\Gamma}}\dim\sL_{\Lambda}\,\mathrm{d}\hat{m}(\Lambda)
  = 1.
\end{displaymath}
The following rule satisfied by the Fourier transformation is also of
crucial importance:
\begin{displaymath}
  \forall s\in\Gamma,\forall f\in L^{2}(\Gamma),\,\,
  \sF[f(s\cdot g)](\Lambda) = \Lambda(s^{-1})\sF[f(g)](\Lambda).
\end{displaymath}

Now we are going to construct a unitary mapping
\[
\Phi:L^{2}(\tilde{M})\to\int_{\hat{\Gamma}}^{\oplus}
\sL_{\Lambda}^{\ast}\otimes\sH_{\Lambda}\,\mathrm{d}\hat{m}(\Lambda)
\]
which makes it possible to decompose the Hamiltonian $H$. Observe that
the tensor product $\sL_{\Lambda}^{*}\otimes\sH_{\Lambda}$ can be
naturally identified with the Hilbert space of
$1\otimes\Lambda$-equivariant operator-valued functions on $\tilde{M}$
with values in
$\sL_{\Lambda}^{*}\otimes\sL_{\Lambda}\equiv\sI_{2}(\sL_{\Lambda})$.
For $f\in L^{2}(\tilde{M})$ and $y\in\tilde{M}$ set
\[
\forall s\in\Gamma,\, f_{y}(s)=f(s^{-1}\cdot y).
\]
The norm $\|f_{y}\|$ in $L^{2}(\Gamma)$ is a $\Gamma$-invariant
function of $y\in\tilde{M}$, and the projection of this function onto
$M$ can be checked to be square integrable. Hence for almost all $x\in
M$ and all $y\in\pi^{-1}(\{ x\})$ it holds $f_{y}\in L^{2}(\Gamma)$.
We define the component $\Phi[f](\Lambda)$, $\Lambda\in\hat{\Gamma}$,
by the prescription
\begin{displaymath}
  \Phi[f](\Lambda)\,(y)
  := \sF[f_{y}](\Lambda)\in\sI_{2}(\sL_{\Lambda}).
\end{displaymath}
In particular, if $f\in L^{1}(\tilde{M})\cap L^{2}(\tilde{M})$ then
\[
\Phi[f](\Lambda)\,(y)=\sum_{s\in\Gamma}f(s^{-1}\cdot y)\Lambda(s).
\]

Equivalently one can define $\Phi$ in the following way. For
$\varphi\in C_{0}^{\infty}(\tilde{M})$, $v\in\sL_{\Lambda}$ and
$y\in\tilde{M}$ set
\begin{equation}
  \Phi[\varphi](\Lambda)(y)v
  = \left(\Phi_{\Lambda}\,\varphi\otimes v\right)\!(y).
  \label{eq:Phi_PhiL}
\end{equation}
Then $\Phi$ introduced in (\ref{eq:Phi_PhiL}) is an isometry and
extends unambiguously to a unitary mapping.

Finally one can verify the formula
\begin{displaymath}
  \Phi H\Phi^{-1} = \int_{\hat{\Gamma}}^{\oplus}1
  \otimes H_{\Lambda}\,\mathrm{d}\hat{m}(\Lambda)
\end{displaymath}
which represents the sought Bloch decomposition. As a corollary we
have
\begin{equation}
  \Phi U(t)\Phi^{-1} = \int_{\hat{\Gamma}}^{\oplus}1
  \otimes U_{\Lambda}(t)\,\mathrm{d}\hat{m}(\Lambda).
  \label{eq:U_int_UL}
\end{equation}

\subsection{A construction for propagators associated to periodic\\
  Hamiltonians}
\label{sec:propag_periodic}

In equality (\ref{eq:U_int_UL}), the evolution operator $U(t)$ is
expressed in terms of $U_{\Lambda}(t)$, $\Lambda\in\hat{\Gamma}$.  It
is possible to invert this relationship and to derive a formula for
the propagator associated to $H_{\Lambda}$ which is expressed in terms
of the propagator associated to $H$.

The propagators are regarded as distributions which are introduced as
kernels of the corresponding evolution operators. Recall that by the
Schwartz kernel theorem (see, for example,
\cite[Theorem~5.2.1]{hoermander}), to every
$B\in\sB(L^{2}(\tilde{M}))$ there exists one and only one
$\beta\in\sD'(\tilde{M}\times\tilde{M})$ such that
\[
\forall\varphi_{1},\varphi_{2}\in C_{0}^{\infty}(\tilde{M}),\quad
\beta(\overline{\varphi_{1}}\otimes\varphi_{2})
= \langle\varphi_{1},B\varphi_{2}\rangle.
\]
Moreover, the map $B\mapsto\beta$ is injective. One calls $\beta$ the
kernel of $B$.

The kernel theorem can be extended to Hilbert spaces formed by
$\Lambda$-equivariant vector-valued functions. In this case the
kernels are operator-valued distributions. To every
$B\in\sB(\sH_{\Lambda})$ there exists one and only one
$\beta\in\sD'(\tilde{M}\times\tilde{M})\otimes\sB(\sL_{\Lambda})$ such
that
\begin{eqnarray*}
  &  & \forall\varphi_{1},\varphi_{2}\in C_{0}^{\infty}(\tilde{M}),
  \forall v_{1},v_{2}\in\sL_{\Lambda},\\
  &  & \langle v_{1},\beta(\overline{\varphi_{1}}
  \otimes\varphi_{2})v_{2}\rangle
  = \langle\Phi_{\Lambda}\,\varphi_{1}
  \otimes v_{1},B\,\Phi_{\Lambda}\,\varphi_{2}
  \otimes v_{2}\rangle.
\end{eqnarray*}
The distribution $\beta$ is $\Lambda$-equivariant:
\begin{displaymath}
  \forall s\in\Gamma,\quad\beta(s\cdot y_1,y_2)
  = \Lambda(s)\beta(y_1,y_2),\,\,\beta(y_1,s\cdot y_2)
  = \beta(y_1,y_2)\Lambda(s^{-1})
\end{displaymath}
In this case, too, the map $B\mapsto\beta$ is injective.

Denote by $\mathcal{K}_{t}\in\sD'(\tilde{M}\times\tilde{M})$ the
kernel of $U(t)\in\sB(L^{2}(\tilde{M}))$, and by
$\mathcal{K}_{t}^{\Lambda}\in\sD'(\tilde{M}\times\tilde{M})
\otimes\sB(\sL_{\Lambda})$
the kernel of $U_{\Lambda}(t)\in\sB(\sH_{\Lambda})$. Here and
everywhere in this section, $t$ is a real parameter. The kernel
$\mathcal{K}_{t}^{\Lambda}$ is $\Lambda$-equivariant:
\begin{displaymath}
  \forall s\in\Gamma,\quad\mathcal{K}_{t}^{\Lambda}(s\cdot y_{1},y_{2})
  = \Lambda(s)\mathcal{K}_{t}^{\Lambda}(y_{1},y_{2}),\,
  \mathcal{K}_{t}^{\Lambda}(y_{1},s\cdot y_{2})
  = \mathcal{K}_{t}^{\Lambda}(y_{1},y_{2})\Lambda(s^{-1}).
\end{displaymath}

First we rewrite the Bloch decomposition of the propagator
(\ref{eq:U_int_UL}) in terms of kernels. It is possible to prove that,
for all $\varphi_{1},\varphi_{2}\in C_{0}^{\infty}(\tilde{M})$, the
function
$\Lambda\mapsto\Tr[\mathcal{K}_{t}^{\Lambda}
(\overline{\varphi_{1}}\otimes\varphi_{2})]$
is integrable on $\hat{\Gamma}$ and
\begin{displaymath}
  \mathcal{K}_{t}(\varphi_{1}\otimes\varphi_{2})
  = \int_{\hat{\Gamma}}\Tr[\mathcal{K}_{t}^{\Lambda}
  (\varphi_{1}\otimes\varphi_{2})]\,\mathrm{d}\hat{m}(\Lambda).
\end{displaymath}
An inverse relation was derived by Schulman in the framework of path
integration \cite{schulman1,schulman2} and reads
\begin{equation}
  \mathcal{K}_{t}^{\Lambda}(x,y)
  = \sum_{s\in\Gamma}\Lambda(s)\,\mathcal{K}_{t}(s^{-1}\cdot x,y).
  \label{eq:schulman_ansatz}
\end{equation}

It is possible to give (\ref{eq:schulman_ansatz}) the following
rigorous interpretation.  Suppose that $\varphi_{1},\varphi_{2}\in
C_{0}^{\infty}(\tilde{M})$ are fixed but otherwise arbitrary. Set
\begin{displaymath}
  F_{t}(s) = \mathcal{K}_{t}\!
  \left(\varphi_{1}(s^{-1}\cdot y_1)\otimes\varphi_{2}(y_2)\right)\,
  \textrm{for~}s\in\Gamma,
\end{displaymath}
and
\begin{displaymath}
  G_{t}(\Lambda) = \mathcal{K}_{t}^{\Lambda}
  (\varphi_{1}\otimes\varphi_{2})\in\sI_{2}(\sL_{\Lambda})\,\,
  \textrm{for~}\Lambda\in\hat{\Gamma}.
\end{displaymath}
One can show that $F_{t}\in L^{2}(\Gamma)$ and $G_{t}$ is bounded on
$\hat{\Gamma}$ in the Hilbert-Schmidt norm. Recalling that
$\hat{m}(\hat{\Gamma})\leq1$ we have
$\|G_{t}(\cdot)\|\in{}L^{1}(\hat{\Gamma})\cap L^{2}(\hat{\Gamma})$. In
\cite{kocabovastovicek} it is verified that
\[
F_{t}=\sF^{-1}[G_{t}].
\]
and, consequently,
\begin{equation}
  G_{t}=\sF[F_{t}].
  \label{schulman}
\end{equation}
Rewriting (\ref{schulman}) formally gives equality
(\ref{eq:schulman_ansatz}).

\section{The Aharonov-Bohm effect with two vortices: the propagator on
  the universal covering space}
\label{sec:ABpropag_cover}

\subsection{A formula for the propagator}

The configuration space for the Aharonov-Bohm effect with two vortices
is the plane with two excluded points, $M=\R^2\setminus\{a,b\}$. This
is a flat Riemannian manifold and the same is true for the universal
covering space $\tilde{M}$. Let $\pi:\tilde{M}\to{}M$ be the
projection. It is convenient to complete the manifold $\tilde{M}$ by a
countable set of points $\AA\cup\BB$ which lie on the border of
$\tilde{M}$ and project onto the excluded points, $\pi(\AA)=\{a\}$ and
$\pi(\BB)=\{b\}$.

$\tilde{M}$ looks locally like $\R^2$ but differs from the Euclidean
space by some global features. First of all, not every two points from
$\tilde{M}$ can be connected by a geodesic segment. Fix a point
$x\in\tilde{M}$. The symbol $D(x)$, as introduced below in
(\ref{eq:Dx_def}), stands for the set of points $y\in\tilde{M}$ which
can be connected with $x$ by a geodesic segment. The domain $D(x)$ is
one sheet of the covering $\tilde{M}\to{}M$. It can be identified with
$\R^2$ cut along two halflines with the limit points $a$ and $b$,
respectively. Thus the border $\partial{}D(x)$ is formed by four
halflines. The universal covering space $\tilde{M}$ can be imagined as
a result of an infinite process of glueing together countably many
copies of $D(x)$ with each copy having four neighbors.

The fundamental group of $M$, called $\Gamma$, is known to be the free
group with two generators $g_a$ and $g_b$. For the generator $g_a$ one
can choose the homotopy class of a simple positively oriented loop
winding once around the point $a$ and leaving the point $b$ in the
exterior. Analogously one can choose the generator $g_b$ by
interchanging the role of $a$ and $b$. One-dimensional unitary
representations $\Lambda$ of $\Gamma$ are determined by two numbers
$\alpha$, $\beta$, $0\leq\alpha,\beta<1$, such that
\begin{displaymath}
  \Lambda(g_a) = e^{2\pi i\alpha},\textrm{~}
  \Lambda(g_b) = e^{2\pi i\beta}.
\end{displaymath}

The standard way to define the Aharonov-Bohm Hamiltonian with two
vortices is to choose a vector potential $\overrightarrow{A}$ for
which $\rot\overrightarrow{A}=0$ on $M$ and such that the
nonintegrable phase factor \cite{wuyang} for a closed path from the
homotopy class $g_a$ or $g_b$ equals $e^{2\pi{}i\alpha}$ or
$e^{2\pi{}i\beta}$, respectively (assuming that $0<\alpha,\beta<1$).
The Hamiltonian then acts as the differential operator
$(-i\nabla-\overrightarrow{A})^2$ in $L^2(M)$. A unitarily equivalent
and for our purposes more convenient possibility is to work with the
Hamiltonian $H_\Lambda=-\Delta$ in the Hilbert space $\sH_\Lambda$ of
$\Lambda$-equivariant functions on $\tilde{M}$, as introduced in
Section~\ref{sec:periodic_H}. Parallelly one considers the free
Hamiltonian $H=-\Delta$ in $L^2(\tilde{M})$. $H$ is $\Gamma$-periodic.
In order to compute, according to prescription
(\ref{eq:schulman_ansatz}), the propagator $\KK^\Lambda(t,x,y)$
associated to $H_\Lambda$ one needs to derive a formula for the free
propagator $\KK(t,x,y)$ on $\tilde{M}$. Such a formula is recalled
below following \cite{pla89}.

Let $\vartheta$ be the Heaviside step function. For
$x,y\in\tilde{M}\cup\AA\cup\BB$ set $\chi(x,y)=1$ if the points $x$,
$y$ can be connected by a geodesic segment, and $\chi(x,y)=0$
otherwise. Given in addition $t\in\R$ we define
\begin{displaymath}
  Z(t,x,y) = \vartheta(t)\chi(x,y)\,
  \frac{1}{4\pi it}\,
  \exp\!\left(\frac{i}{4t}\,\dist^{2}(x,y)\right),\\
\end{displaymath}
Furthermore, for $x_1,x_2,x_3\in\tilde{M}\cup\AA\cup\BB$ such that
$\chi(x_1,x_2)=\chi(x_2,x_3)=1$, and for $t_1,t_2>0$ we set
\begin{displaymath}
  V\!\left(
    \begin{array}{c}
      x_{3},x_{2},x_{1}\\
      t_{2},t_{1}
    \end{array}
  \right) = 2i\left(\left(\theta-\pi
      +i\log\!\left(\frac{t_{2}r_{1}}{t_{1}r_{2}}\right)\right)^{\!-1}
    -\left(\theta+\pi
      +i\log\!\left(\frac{t_{2}r_{1}}{t_{1}r_{2}}\right)\right)^{\!-1}
  \right)
\end{displaymath}
where $\theta=\angle\,x_{1},x_{2},x_{3}\in\R$ is the oriented angle
and $r_{1}=\dist(x_{1},x_{2})$, $r_{2}=\dist(x_{2},x_{3})$. Note that
if the inner vertex $x_2$ belongs to the set of extreme points
$\AA\cup\BB$ then the angle $\theta$ can take any real value.

We claim that the free propagator on $\tilde{M}$ equals
\begin{equation}
  \label{eq:free_propag_sum}
  \KK(t,x,x_0) = \sum_{\gamma}\KK_\gamma(t,x,x_0)
\end{equation}
where the sum runs over all piecewise geodesic curves
$\gamma:x_0\to{}C_1\to\ldots\to{}C_n\to{}x$ with the inner vertices
$C_j$, $1\leq{}j\leq{}n$, belonging to the set of extreme points
$\AA\cup\BB$. This means that it should hold
$\chi(x_0,C_1)=\chi(C_1,C_2)=\ldots=\chi(C_n,x)=1$. Let us denote by
$|\gamma|=n$ the length of the sequence $(C_{1},C_{2},\ldots,C_{n})$.
In particular, if $|\gamma|=0$ then $\gamma$ designates the geodesic
segment $x_0\to{}x$. To simplify notation we set everywhere where
convenient $C_0=x_0$ and $C_{n+1}=x$. With this convention, the
summands in (\ref{eq:free_propag_sum}) equal
\begin{eqnarray}
  & & \KK_{\gamma}(t,x,x_{0})\nonumber \\
  & & = \,\int_{\R^{n+1}}\mathrm{d}t_{n}\ldots\mathrm{d}t_{0}\,
  \delta(t_{n}+\ldots+t_{0}-t)\,
  \prod_{j=0}^{n-1} V\!\left(
    \begin{array}{c}
      C_{j+2},C_{j+1},C_{j}\\
      t_{j+1},t_{j}
    \end{array}
  \right)\prod_{j=0}^{n} Z(t_{j},C_{j+1},C_{j}). \nonumber\\
  \label{eq:free_propag_Kgamma}
\end{eqnarray}
In particular, if $|\gamma|=0$ then
$\KK_{\gamma}(t,x,x_{0})=Z(t,x,x_{0})$, and if $|\gamma|=1$ then
$\gamma$ designates a path composed of two geodesic segments
$x_0\to{}C\to{}x$, with $C\in\AA\cup\BB$, and
\begin{displaymath}
  \KK_{\gamma}(t,x,x_{0}) = \vartheta(t)\int_0^t\,
  V\!\left(
    \begin{array}{c}
      x,C,x_{0}\\
      t-s,s
    \end{array}
  \right)Z(t-s,x,C)Z(s,C,x_{0})\,\mathrm{d}s\,.
\end{displaymath}

\subsection{Auxiliary relations}

As is well known, in $\R^2$ it holds true that
\begin{displaymath}
  \left(\frac{\partial}{\partial x}
    +i\,\frac{\partial}{\partial y}\right)\frac{1}{x+i y}
  = 2\pi\delta(x)\delta(y)
\end{displaymath}
and, consequently,
\begin{displaymath}
  \Delta\,\frac{1}{x+i y}
  = 2\pi\left(\delta(y)\delta'(x)-i\,\delta(x)\delta'(y)\right).
\end{displaymath}
With the aid of the last equality one can verify that the relation
\begin{equation}
  \label{eq:auxrel1}
   \left(\frac{\partial^2}{\partial r^2}
     +\frac{1}{r}\,\frac{\partial}{\partial r}
     +\frac{1}{r^2}\,\frac{\partial^2}{\partial \theta^2}\right)
   \left(\theta+i\log\!\left(\frac{t}{r}\right)\right)^{\!-1}
   = \frac{2\pi t}{r^2}
   \left(\delta(t-r)\delta'(\theta)
     -i r\delta'(t-r)\delta(\theta)\right)
\end{equation}
is valid on the domain $t>0$, $r>0$, $\theta\in\R$. It is
straightforward to see that, on the same domain,
\begin{equation}
  \label{eq:auxrel2}
  \left(r\,\frac{\partial}{\partial r}
    +t\,\frac{\partial}{\partial t}\right)
  \left(\theta+i\log\!\left(\frac{t}{r}\right)\right)^{\!-1}
  = 0.
\end{equation}
Combining (\ref{eq:auxrel1}), (\ref{eq:auxrel2}) and the Leibniz rule
one finds that
\begin{eqnarray}
  && \left(i\,\frac{\partial}{\partial t}
    +\frac{\partial^2}{\partial r^2}
      +\frac{1}{r}\,\frac{\partial}{\partial r}+\frac{1}{r^2}\,
      \frac{\partial^2}{\partial\theta^2}\right)
  \left(\theta+i\log\!\left(\frac{t}{r}\right)\right)^{\!-1}
  \frac{1}{t}\,\exp\!\left(i\,\frac{r^2}{4t}\right) \nonumber\\
  \label{eq:auxrel3}
  && = \frac{2\pi}{r^2}\,\exp\!\left(i\,\frac{r^2}{4t}\right)
  \left(\delta(t-r)\delta'(\theta)
    -i r\delta'(t-r)\delta(\theta)\right).
\end{eqnarray}
Equipped with (\ref{eq:auxrel3}) one can prove the equality
\begin{eqnarray}
  & & \left(i\,\frac{\partial}{\partial t}
    +\frac{\partial^2}{\partial r^2}
      +\frac{1}{r}\,\frac{\partial}{\partial r}+\frac{1}{r^2}\,
      \frac{\partial^2}{\partial\theta^2}\right)
  \int_{0}^{t}\left(\theta
    +i\log\!\left(\frac{(t-s)r_{0}}{sr}\right)\right)^{\!-1} 
  \nonumber\\
  & & \qquad\qquad\qquad\qquad\qquad\qquad\qquad\quad
  \times\,\frac{1}{t-s}\,
  \exp\!\left(i\,\frac{r^{2}}{4(t-s)}\right)f(s)\,\mathrm{d}s
  \nonumber\\
  & & =\,\frac{2\pi r_{0}}{r^{2}(r+r_{0})}\,
  \exp\!\left(i\,\frac{(r+r_{0})r}{4t}\right)
  \Bigg[f\!\left(\frac{tr_{0}}{r+r_{0}}\right)\delta'(\theta)
  \nonumber\\
  & & \quad-\,i\,\frac{r}{r+r_{0}}
  \left(\left(1+i\,\frac{r_{0}(r+r_{0})}{4t}\right)
    f\!\left(\frac{tr_{0}}{r+r_{0}}\right)+\frac{tr_{0}}{r+r_{0}}\,
    f'\!\left(\frac{tr_{0}}{r+r_{0}}\right)\right)\delta(\theta)
  \Bigg] \nonumber\\
  \label{eq:auxrel_main}
\end{eqnarray}
which is true in the sense of distributions for any $r_0>0$ and
$f\in{}C^1([0,+\infty[\,)$, again on the domain $t>0$, $r>0$,
$\theta\in\R$. Note that
\begin{displaymath}
  \frac{1}{\varepsilon}\,
  \exp\!\left({i\,\frac{r^2}{4\varepsilon}}\right)\to 0
  \textrm{~~as~}\varepsilon\to 0+
  \textrm{~in~}\sD'(\,]0,+\infty[\,).
\end{displaymath}
In particular, letting
\begin{displaymath}
  f(s) = \frac{1}{s}\exp\!\left(i\,\frac{r_0^{\,2}}{4s}\right)
\end{displaymath}
one derives the following equality which is true in the sense of
distributions,
\begin{eqnarray}
  & & \left(i\,\frac{\partial}{\partial t}
    +\frac{\partial^2}{\partial r^2}
      +\frac{1}{r}\,\frac{\partial}{\partial r}+\frac{1}{r^2}\,
      \frac{\partial^2}{\partial\theta^2}\right)
  \int_{0}^{t}\left(\theta
    +i\log\!\left(\frac{(t-s)r_{0}}{sr}\right)\right)^{\!-1} 
  \nonumber\\
  & & \qquad\qquad\qquad\qquad\qquad\qquad\quad
  \times\,\frac{1}{(t-s)s}\,
  \exp\!\left(i\left(\frac{r^{2}}{4(t-s)}+\frac{r_0^{\,2}}{4s}\right)
  \right)\,\mathrm{d}s
  \nonumber\\
  & & =\,\frac{2\pi}{tr^{2}}\,
  \exp\!\left(i\,\frac{(r+r_{0})^2}{4t}\right)\delta'(\theta).
  \label{eq:auxrel_exp}
\end{eqnarray}

Let us also recall the following basic fact concerning the generalized
Laplacian. If $G\subset\tilde{M}$ is an open set with a piecewise
smooth boundary, $\chi_G$ is the characteristic function of $G$,
$\overrightarrow{n}$ is the normalized outer normal vector field on
$\partial{}G$ and $\eta$ is a smooth function on $\tilde{M}$ then, in
the sense of distributions,
\begin{equation}
  \label{eq:generalized_laplacian}
  \Delta(\eta\,\chi_G) = (\Delta\eta)\chi_G
  -\frac{\partial\eta}{\partial\overrightarrow{n}}\,
  \delta_{\partial G}
  -\frac{\partial}{\partial\overrightarrow{n}}
  (\eta\,\delta_{\partial G}).
\end{equation}
The distribution $\delta_{\partial{}G}$ is the layer supported on the
curve $\partial{}G$ which is defined by the curve integral
\begin{displaymath}
  \forall\varphi\in C_0^\infty(\tilde{M}),\textrm{~}
  \delta_{\partial G}(\varphi)
  = \int_{\partial G}\varphi\,\dd\ell.
\end{displaymath}

\subsection{Verification of the formula}

We have to show that, for $x_0\in\tilde{M}$ fixed, the propagator
$\KK(t,x,x_0)$ defined in (\ref{eq:free_propag_sum}),
(\ref{eq:free_propag_Kgamma}) verifies the condition
\begin{displaymath}
  \left(i\frac{\partial}{\partial t}+\Delta\right)
  \KK(t,x,x_{0}) = i\,\delta(t)\delta(x,x_{0})
  \textrm{~~on~}\R\times\tilde{M}.
\end{displaymath}
This is equivalent to showing that
\begin{equation}
  \label{eq:Kt_limit_t0}
  \lim_{t\rightarrow0_{+}}\KK(t,x,x_{0}) = \delta(x,x_{0})
\end{equation}
and
\begin{equation}
  \label{eq:schroedeq_Kt_0}
  \left(i\frac{\partial}{\partial t}+\Delta\right)
  \KK(t,x,x_{0}) = 0
  \textrm{~~for~}t>0,x\in\tilde{M}.
\end{equation}

Equality (\ref{eq:Kt_limit_t0}) is rather obvious. Since $Z(t,x,x_0)$
looks on the sheet $\{x;\,\chi(x,x_0)=1\}$ as the free propagator on
$\R^2$ we have
\begin{displaymath}
  \lim_{t\rightarrow0_{+}}Z(t,x,x_{0}) = \delta(x,x_{0}).
\end{displaymath}
From a similar reason, $\lim_{t\rightarrow0_{+}}Z(t,x,C)=0$ if
$C\in\AA\cup\BB$ and $x$ runs over $\tilde{M}$. Hence
\begin{displaymath}
  \lim_{t\rightarrow0_{+}}\KK_\gamma(t,x,x_{0}) = 0
  \textrm{~~if~}|\gamma|\geq1.
\end{displaymath}

Let us proceed to the verification of (\ref{eq:schroedeq_Kt_0}).
First we introduce some notation related to the geometry of the
universal covering space $\tilde{M}$. Denote by $\varrho$ the distance
$\dist(a,b)$. Observe that if $C_1,C_2\in\AA\cup\BB$ then
$\chi(C_1,C_2)=1$ if and only if $\dist(C_1,C_2)=\varrho$. If this is
the case then necessarily $C_1\in\AA$ and $C_2\in\BB$ or vice versa.

For $x\in\tilde{M}\cup\AA\cup\BB$ set
\begin{equation}
  \label{eq:Dx_def}
  D(x) = \left\{y\in\tilde{M};\,\chi(x,y)=1\right\}.
\end{equation}
If $x\in\tilde{M}$ then $D(x)$ can be identified with the aid of the
projection $\pi:\tilde{M}\to{}M$ with the plane cut along two
halflines with the limit points $a$ and $b$, respectively. Thus the
border of $D(x)$ consists of two pairs of halflines. One pair has a
common limit point $A\in\AA$ and is denoted $\partial{}D(x;A)$, the
other pair has a common limit point $B\in\BB$ and is denoted
$\partial{}D(x;B)$. We have
\begin{equation}
  \label{eq:border_Dx}
  \partial D(x) = \partial D(x;A)\cup\partial D(x;B).
\end{equation}

If $C\in\AA\cup\BB$ then $D(C)$ resembles the universal covering space
in the one-vortex case. It can be viewed as a union of countably many
sheets glued together in a staircase-like way. Each sheet contributes
to the border of $D(C)$ by a pair of halflines with a common limit
point $C'$. Thus the border $\partial{}D(C)$ is formed by a countable
union of pairs of halflines: if $C\in\AA$ then we write
\begin{equation}
  \label{eq:border_DCA}
  \partial D(C) = \bigcup_{C'\in\BB,\,\dist(C,C')=\varrho}
  \partial D(C;C'),
\end{equation}
if $C\in\BB$ then
\begin{equation}
  \label{eq:border_DCB}
  \partial D(C) = \bigcup_{C'\in\AA,\,\dist(C,C')=\varrho}
  \partial D(C;C').
\end{equation}

Let us first examine the case $|\gamma|=0$. It holds
\begin{displaymath}
  \left(i\frac{\partial}{\partial t}+\Delta\right)Z(t,x,x_{0}) = 0.
\end{displaymath}
for $t>0$ and $x\in{}D(x_{0})$. Observe also that
\begin{displaymath}
  \frac{\partial}{\partial\overrightarrow{n}}\,Z(t,x,x_{0}) = 0
  \textrm{~~for~}x\in\partial D(x_{0})
\end{displaymath}
where $\overrightarrow{n}$ is the normalized outer normal vector field
on the border $\partial{}D(x_{0})$. This is so since, in the polar
coordinates centered at $x_0$, $Z(t,x,x_0)$ does not depend on the
angle variable. Let us also note that the function $Z(t,x,x_0)$ can be
continued smoothly in the variable x over the borderline of the domain
$D(x_0)$. Thus, in virtue of (\ref{eq:generalized_laplacian}), we
obtain (for $t>0$, $x\in\tilde{M}$)
\begin{equation}
  \label{eq:gamma0}
  \left(i\frac{\partial}{\partial t}+\Delta\right)Z(t,x,x_{0})
  = -\frac{\partial}{\partial\overrightarrow{n}}
  \left(Z(t,x,x_{0})\,\delta_{\partial D(x_{0})}\right).
\end{equation}

\begin{rem*}
  In (\ref{eq:gamma0}) as well as everywhere in this section we use
  the following convention. The value of a density (which is in this
  case $Z(t,x,x_0)$) on the border $\partial{}D(x_0)$ is understood as
  the limit achieved from the interior of the domain $D(x_0)$.
\end{rem*}

Next we discuss the case $|\gamma|=1$. Then $\gamma$ designates a
piecewise geodesic curve $x_0\to{}C\to{}x$, with $C\in\AA\cup\BB$.
Denote by $\gamma'$ the geodesic segment $x_0\to{}x$ (provided
$x\in{}D(x_0)$). We have
\begin{eqnarray}
  && \KK_\gamma(t,x,x_{0})
  \,=\, \frac{1}{8\pi^{2}i}\,\chi(x,C)\chi(C,x_{0})
  \nonumber\\
  &  & \qquad\times\,\int_{0}^{t}\left(
    \left(\theta-\pi
      +i\log\!\left(\frac{(t-s)r_{0}}{sr}\right)\right)^{\!-1}
    -\left(\theta+\pi
      +i\log\!\left(\frac{(t-s)r_{0}}{sr}\right)\right)^{\!-1}
  \right)\nonumber\\
  \label{eq:contrib_gamma1}
  &  & \qquad\qquad\times\,\frac{1}{(t-s)s}\,
  \exp\!\left(i
    \left(\frac{r^{2}}{4(t-s)}+\frac{r_{0}^{2}}{4s}\right)\right)ds
\end{eqnarray}
where $r=\dist(x,C)$, $r_{0}=\dist(C,x_{0})$ and
$\theta=\angle\,x_{0},C,x$.

An application of the differential operator $(i\partial_t+\Delta)$ to
the RHS of (\ref{eq:contrib_gamma1}) in the sense of distributions
results in several singular terms supported on one-dimensional
submanifolds. First, due to the discontinuity of the characteristic
function $\chi(x,C)$, the application of the Laplace operator leads to
two terms supported on the boundary $\partial{}D(C)$ (see
(\ref{eq:generalized_laplacian})). Second, as it follows from
(\ref{eq:auxrel_exp}), the singularity of the integrand for the values
$\theta=\pm\pi$ and $r_0/s=r/(t-s)$ produces terms supported on the
submanifold determined by $\theta=\pm\pi$, and this set is nothing but
a part of the boundary of the domain $D(x_0)$, namely
$\partial{}D(x_0;C)$. Notice that for $\theta=\pm\pi$ it holds
$r+r_{0}=\dist(x,x_{0})$ and
$\partial/\partial\overrightarrow{n}=\pm{}r^{-1}\partial/\partial\theta$.
Moreover, in the polar coordinates centered at $C$,
\begin{displaymath}
  \delta_{\partial D(x_{0};C)}
  = \frac{1}{r}\left(\delta(\theta-\pi)+\delta(\theta+\pi)\right).
\end{displaymath}
Thus the latter contribution takes the form
\begin{eqnarray*}
  \frac{1}{4\pi itr^{2}}
  \frac{\partial}{\partial\theta}
  \left(\exp\!\left(\frac{i}{4t}\,\dist(x,x_{0})^{2}\right)
    (\delta(\theta-\pi)-\delta(\theta+\pi))\right)
  = \frac{\partial}{\partial\overrightarrow{n}}
  \left(\KK_{t}^{\gamma'}(x,x_{0})\delta_{\partial D(x_{0},C)}\right)
\end{eqnarray*}
where $\KK_{\gamma'}(t,x,x_{0})=Z(t,x,x_{0})$. In summary, we obtain
\begin{eqnarray}
  \left(i\frac{\partial}{\partial t}+\Delta\right)
  \KK_{\gamma}(t,x,x_{0})
  & = &
  -\,\left(\frac{\partial}{\partial\overrightarrow{n}}\,
    \KK_{\gamma}(t,x,x_{0})\right)\delta_{\partial D(G)}
  -\frac{\partial}{\partial\overrightarrow{n}}
  \left(\KK_{\gamma}(t,x,x_{0})\delta_{\partial D(C)}\right)
  \nonumber\\
  \label{eq:gamma1}
  & & +\,\frac{\partial}{\partial\overrightarrow{n}}
  \left(\KK_{\gamma'}(t,x,x_{0})\delta_{\partial D(x_{0},C)}\right).
\end{eqnarray}

Finally let us consider the case $|\gamma|\geq2$. Thus $\gamma$ is a
piecewise geodesic curve $x_0\to{}C_{1}\to\ldots\to{}C_{n}\to{}x$,
$n\geq2$. Denote by $\gamma'$ the truncated geodesic curve
$x_0\to{}C_{1}\to\ldots\to{}C_{n-1}\to{}x$ (provided
$x\in{}D(C_{n-1})$). One can express
\begin{eqnarray}
  \KK_{\gamma}(t,x,x_{0})
  & =  & \int_{\R^{n}}\mathrm{d}t_{n-1}\ldots\mathrm{d}t_{0}\,
  V\!\left(
    \begin{array}{c}
      x,C_{n},C_{n-1}\\
      t-\tau,t_{n-1}
    \end{array}
  \right)Z(t-\tau,x,C_{n})F_{\gamma}(t_{0},\ldots,t_{n-1},x_{0})
  \nonumber\\
  & = & \frac{1}{2\pi}\,\chi(x,C_{n})
  \int_{\R^{n-1}}\mathrm{d}t_{n-2}\ldots\mathrm{d}t_{0}
  \int_0^{t-\tau'}\dd t_{n-1} \nonumber\\
  &  & \times\,\left(\frac{1}{\theta-\pi+i
      \log\!\left(\frac{(t-\tau'-t_{n-1})\varrho}{t_{n-1}r}\right)}
    -\frac{1}{\theta+\pi+i
      \log\!\left(\frac{(t-\tau'-t_{n-1})\varrho}{t_{n-1}r}\right)}
  \right)\nonumber\\
  \label{eq:contrib_gamma_geq2}
  &  & \times\,\frac{1}{t-\tau'-t_{n-1}}\,
  \exp\!\left(i\,\frac{r^{2}}{4(t-\tau'-t_{n-1})}\right)
  F_{\gamma}(t_{0},\ldots,t_{n-1},x_{0})
\end{eqnarray}
where
\begin{displaymath}
  \tau = t_{0}+\ldots+t_{n-2}+t_{n-1},\textrm{~}
  \tau' = t_{0}+\ldots+t_{n-2},\textrm{~}r = \dist(C_{n},x),
  \textrm{~}\theta = \angle\,C_{n-1},C_{n},x,
\end{displaymath}
and
\begin{displaymath}
  F_{\gamma}(t_{0},\ldots,t_{n-1},x_{0})
  = \prod_{j=0}^{n-2}V\!\left(
    \begin{array}{c}
      C_{j+2},C_{j+1},C_{j}\\
      t_{j+1},t_{j}
    \end{array}
  \right)\prod_{j=0}^{n-1}Z_{t_{j}}(C_{j+1},C_{j}).
\end{displaymath}

An application of the differential operator $(i\partial_t+\Delta)$ to
the RHS of (\ref{eq:contrib_gamma_geq2}) in the sense of distributions
again produces several singular terms. In consequence of the
discontinuity of the characteristic function $\chi(x,C_n)$ a single
and a double layer supported on the boundary $\partial{}D(C_n)$ occur
(see (\ref{eq:generalized_laplacian})). The singularity of the
integrand for the values $\theta=\pm\pi$ and
$\varrho/t_{n-1}=r/(t-\tau'-t_{n-1})$ produces terms supported on the
part of the boundary of the domain $D(C_{n-1})$, namely on
$\partial{}D(C_{n-1};C_n)$. This time one can apply identity
(\ref{eq:auxrel_main}). In order to treat the resulting terms the
following equalities are useful.

Suppose that $\theta=\pm\pi$ and so $x\in\partial{}D(C_{n-1};C_n)$.
Set
\begin{displaymath}
  r' = r+\varrho = \dist(C_{n-1},x),\textrm{~}
  \theta' = \angle\,C_{n-2},C_{n-1},x.
\end{displaymath}
If $\varrho/t_{n-1}=r/(t-\tau'-t_{n-1})$ then
\begin{displaymath}
  t_{n-1} = \frac{\varrho(t-\tau')}{r'}\textrm{~~and~~}
  \frac{t-\tau'-t_{n-1}}{r} = \frac{t-\tau'}{r'}\,.
\end{displaymath}
Moreover,
\begin{displaymath}
  \frac{\varrho}{r'}\,\exp\!\left(\frac{ir^{2}}{4(t-\tau)}\right)
  Z(t_{n-1},C_n,C_{n-1}) = Z(t-\tau',x,C_{n-1})
\end{displaymath}
and
\begin{displaymath}
  V\!\left(
    \begin{array}{c}
      C_{n},C_{n-1},C_{n-2}\\
      \varrho s_{2}/r',s_{1}
    \end{array}
  \right) = V\!\left(
    \begin{array}{c}
      x,C_{n-1},C_{n-2}\\
      s_{2},s_{1}
    \end{array}\right).  
\end{displaymath}
Observe also that
\begin{displaymath}
  \frac{\partial}{\partial s}
  \left(\exp\!\left(\frac{ir^{2}}{4(t-\tau'-s)}\right)
    \exp\left(\frac{i\varrho^{2}}{4s}\right)
  \right)\Bigg|_{s=\varrho(t-\tau')/r'} = 0,
\end{displaymath}
\begin{displaymath}
  \exp\!\left(\frac{ir^{2}}{4(t-\tau'-s)}\right)
  \exp\!\left(\frac{i\varrho^{2}}{4s}
  \right)\Bigg|_{s=\varrho(t-\tau')/r'}
  = \exp\!\left(\frac{i\,{r'}^{\,2}}{4(t-\tau')}\right),
\end{displaymath}
and for $\theta=\pi$,
\begin{displaymath}
  \frac{\partial}{\partial s}\, V\!\left(
    \begin{array}{c}
      C_{n},C_{n-1},C_{n-2}\\
      s,t_{n-2}
    \end{array}
  \right)\Bigg|_{s=\varrho(t-\tau')/r'}
  = \frac{ir'}{\varrho(t-\tau')}
  \frac{\partial}{\partial\theta'}\, V\!\left(
    \begin{array}{c}
      x,C_{n-1},C_{n-2}\\
      t-\tau',t_{n-2}
    \end{array}\right).
\end{displaymath}
A similar relation holds true also for $\theta=-\pi$.

After a bit tedious but quite straightforward manipulations one
arrives at the final equality
\begin{eqnarray}
  \left(i\frac{\partial}{\partial t}
    +\Delta\right)\KK_{\gamma}(t,x,x_{0})
  & = & -\left(\frac{\partial}{\partial\overrightarrow{n}}\,
    \KK_{\gamma}(t,x,x_{0})\right)
  \delta_{\partial D(C_{n})}
  -\frac{\partial}{\partial\overrightarrow{n}}
  \left(\KK_{\gamma}(t,x,x_{0})\delta_{\partial D(C_{n})}\right)
  \nonumber\\
  & & +\left(\frac{\partial}{\partial\overrightarrow{n}}\,
    \KK_{\gamma'}(t,x,x_{0})\right)
  \delta_{\partial D(C_{n-1};C_{n})} \nonumber\\
  \label{eq:gamma_geq2}
  & & +\,\frac{\partial}{\partial\overrightarrow{n}}
  \left(\KK_{\gamma'}(t,x,x_{0})
  \delta_{\partial D(C_{n-1};C_{n})}\right).
\end{eqnarray}

Now we can show equality (\ref{eq:schroedeq_Kt_0}) when taking into
account (\ref{eq:gamma0}), (\ref{eq:gamma1}) and
(\ref{eq:gamma_geq2}). It is true that
\newpage
\begin{eqnarray*}
  &  & \left(i\frac{\partial}{\partial t}+\Delta\right)
  \KK(t,x,x_{0})\\
  &  & \quad=\, \sum_{|\gamma|\geq2}
  \Bigg[-\left(\frac{\partial}{\partial\overrightarrow{n}}\,
    \KK_{\gamma}(t,x,x_{0})\right)
  \delta_{\partial D(C_{n})}
  -\frac{\partial}{\partial\overrightarrow{n}}
  \left(\KK_{\gamma}(t,x,x_{0})\delta_{\partial D(C_{n})}\right)\\
  &  & \qquad\qquad\quad
  +\left(\frac{\partial}{\partial\overrightarrow{n}}\,
    \KK_{\gamma'}(t,x,x_{0})\right)
  \delta_{\partial D(C_{n-1};C_{n})}
  +\frac{\partial}{\partial\overrightarrow{n}}
  \left(\KK_{\gamma'}(t,x,x_{0})\delta_{\partial D(C_{n-1};C_{n})}
    \right)\Bigg]\\
  &  & \qquad+\,\sum_{|\gamma|=1}
  \Bigg[-\left(\frac{\partial}{\partial\overrightarrow{n}}\,
    \KK_{\gamma}(t,x,x_{0})\right)\delta_{\partial D(G)}
  -\frac{\partial}{\partial\overrightarrow{n}}
  \left(\KK_{\gamma}(t,x,x_{0})\delta_{\partial D(C)}\right)\\
  & & \qquad\qquad\qquad
  +\,\frac{\partial}{\partial\overrightarrow{n}}
  \left(Z(t,x,x_{0})\delta_{\partial D(x_{0};C)}\right)
  \Bigg]
  -\frac{\partial}{\partial\overrightarrow{n}}
  \left(Z(t,x,x_{0})\delta_{\partial D(x_{0})}\right)\\
  & & \quad=\,0
\end{eqnarray*}
where we have used (\ref{eq:border_Dx}), (\ref{eq:border_DCA}) and
(\ref{eq:border_DCB}).

\section{The Aharonov-Bohm effect with two vortices: the propagator
  associated to $H_\Lambda$}
\label{sec:ABpropag_Lambda}

Without loss of generality we can suppose that the vortices are
located in the points $a=(0,0)$ and $b=(\varrho,0)$. Let
$(r_{a},\theta _{a})$ be the polar coordinates centered at the point
$a$ and $(r_{b},\theta _{b})$ be the polar coordinates centered at the
point $b$. To express the propagator for $H_\Lambda$ it is convenient
to pass to a unitarily equivalent formulation. Let us cut the plane
along two half-lines,
\begin{displaymath}
  L_{a}=\,]-\infty ,0[\,\times\{0\}\textrm{~and~}
  L_{b}=\,]\varrho ,+\infty[\,\times\{0\}.
\end{displaymath}
The values $\theta_{a}=\pm\pi$ correspond to the two sides of the cut
$L_{a}$, and similarly for $\theta_{b}$ and $L_{b}$. The unitarily
equivalent Hamiltonian $H_\Lambda'$ is formally equal to $-\Delta$ in
$L^2(\R^2,\dd^2x)$ and is determined by the boundary conditions along
the cut,
\begin{eqnarray*}
  &  & \hspace {-3em}\psi (r_{a},\theta_{a}=\pi)
  = e^{2\pi\,i\,\alpha }\psi (r_{a},\theta_{a}=-\pi ),
  \textrm{~}\partial_{r_{a}}\psi (r_{a},\theta _{a}=\pi )
  = e^{2\pi\,i\,\alpha}\partial_{r_{a}}
  \psi(r_a,\theta_{a}=-\pi)\, ,
  \\
  & & \hspace {-3em}\psi(r_{b},\theta_{b}=\pi)
  = e^{2\pi\,i\,\beta}
  \psi(r_{b},\theta _{b}=-\pi),\textrm{~}
  \partial_{r_{b}}\psi (r_{b},\theta _{b}=\pi)
  = e^{2\pi\,i\,\beta}\partial_{r_{b}}
  \psi(r_b,\theta_{b}=-\pi)\,.
\end{eqnarray*}
In addition, one should impose a boundary condition at the vortices,
namely $\psi(a)=\psi(b)=0$.

Let us denote $D=\R^2\setminus(L_a\cup{}L_b)$. Then one can embed
$D\subset\tilde{M}$ as a fundamental domain. We wish to find a formula
for the propagator ${\KK'}^{\Lambda}(t,x,x_0)$ associated to the
Hamiltonian ${H'}_\Lambda$. It can be simply obtained as the
restriction to $D$ of the propagator $\KK^{\Lambda}(t,x,x_0)$
associated to the Hamiltonian $H_\Lambda$. On the other hand, to
construct $\KK^{\Lambda}(t,x,x_0)$ one can apply formula
(\ref{eq:schulman_ansatz}) and the knowledge of the free propagator on
$\tilde{M}$, see (\ref{eq:free_propag_sum}),
(\ref{eq:free_propag_Kgamma}). Thus we get
\begin{equation}
  \label{eq:propag_2AB}
  \KK^{\Lambda}(t,x,x_0) = \sum_{g\in\Gamma}\sum_{\gamma}
  \Lambda(g^{-1})\,\KK_\gamma(t,g\cdot x,x_0).
\quad
\end{equation}

Fix $t>0$ and $x_0,x\in{}D$. One can classify piecewise geodesic paths
in $\tilde{M}$,
\begin{equation}
  \label{eq:covering_gamma}
  \gamma:x_0\to{}C_1\to\ldots\to{}C_n\to{}g\cdot{}x,
\end{equation}
with $C_j\in\AA\cup\BB$ and $g\in\Gamma$, according to their
projections to $M$. Let $\overline{\gamma}$ be a finite alternating
sequence of points $a$ and $b$, i.e.,
$\overline{\gamma}=(c_1,\ldots,c_n)$, $c_j\in\{a,b\}$ and
$c_j\neq{}c_{j+1}$. The empty sequence $\overline{\gamma}=()$ is
admissible.  Relate to $\overline{\gamma}$ a piecewise geodesic path
in $M$, namely $x_0\to{}c_1\to\ldots{}\to{}c_n\to{}x$. Suppose that
this path is covered by a path $\gamma$ in $\tilde{M}$, as given in
(\ref{eq:covering_gamma}).  Then $C_j\in\AA$ iff $c_j=a$ and
$C_j\in\BB$ iff $c_j=b$. Denote the angles
$\angle\,x_0,c_1,c_2=\theta_0$ and $\angle\,c_{n-1},c_n,x=\theta$.
Then the angles in the path $\gamma$ in (\ref{eq:covering_gamma}) take
the values $\angle\,x_0,C_1,C_2=\theta_0+2\pi{}k_1$,
$\angle\,C_{n-1},C_n,g\cdot{}x=\theta+2\pi{}k_n$ and
$\angle\,C_j,C_{j+1},C_{j+2}=2\pi{}k_{j+1}$ for $1\leq{}j\leq{}n-2$
(if $n\geq3$), where $k_1,\ldots,k_n$ are integers. Any values
$k_1,\ldots,k_n\in\Z$ are possible. In that case the representation
$\Lambda$ applied to the group element $g$ occurring in
(\ref{eq:covering_gamma}) takes the value
\begin{displaymath}
  \Lambda(g) = \exp(2\pi i(k_1\sigma_1+\ldots+k_n\sigma_n))
\end{displaymath}
where $\sigma_j\in\{\alpha,\beta\}$ and $\sigma_j=\alpha$ if $c_j=a$,
and $\sigma_j=\beta$ if $c_j=b$.

Using the equalities
\begin{eqnarray*}
  &  & \sum_{k\in\Z}\exp(2\pi\mathrm{i}\alpha k)
  \left(\frac{1}{\theta+2k\pi-\pi+\mathrm{i}s}
    -\frac{1}{\theta+2\pi k+\pi+\mathrm{i}s}\right)\\
  &  & =\,-\sin(\pi\alpha)\int_{-\infty}^{+\infty}\,
  \frac{\exp((\theta+\mathrm{i}s)\tau)}
  {\sin(\pi(\alpha+\mathrm{i}\tau))}\,\mathrm{d}\tau
\end{eqnarray*}
and
\begin{displaymath}
  \int_{-\infty}^{\infty}
  \frac{\exp((\theta+\mathrm{i}s)\tau)}
  {\sin(\pi(\alpha+\mathrm{i}\tau))}\,\mathrm{d}\tau
  = 2\,\frac{\exp(-\alpha(s-\mathrm{i}\theta))}
  {1+\exp(-s+\mathrm{i}\theta)},\,
\end{displaymath}
that are valid for $0<\alpha<1$, $|\theta|<\pi$, one can carry out a
partial summation in (\ref{eq:propag_2AB}) over the integers
$k_1,\ldots,k_n$. This way the double sum in (\ref{eq:propag_2AB})
reduces to a sum over finite alternating sequences
$\overline{\gamma}$.

Let us conclude our contribution by giving the resulting formula for
${\KK'}^{\Lambda}(t,x,x_0)$. We set
\begin{displaymath}
  \zeta _{a}=1\textrm{~or~}\zeta _{a}=e^{2\,\pi\,i\,\alpha}
  \textrm{~or~}\zeta_{a} = e^{-2\,\pi\,i\,\alpha}
\end{displaymath}
depending on whether the segment $\overline{x_{0}x}$ does not intersect
$L_{a}$, or $\overline{x_{0}x}$ intersects $L_{a}$ and $x_{0}$
lies in the lower half-plane, or $\overline{x_{0}x}$ intersects $L_{a}$
and $x_{0}$ lies in the upper half-plane. Analogously,
\begin{displaymath}
  \zeta _{b}=1\textrm{ or }\zeta _{b}=e^{2\,\pi\,i\,\beta}
  \textrm{~or~}\zeta _{b}=e^{-2\,\pi\,i\,\beta}
\end{displaymath}
depending on whether the segment $\overline{x_{0}x}$ does not intersect
$L_{b}$, or $\overline{x_{0}x}$ intersects $L_{b}$ and $x_{0}$
lies in the upper half-plane, or $\overline{x_{0}x}$ intersects $L_{b}$
and $x_{0}$ lies in the lower half-plane. Furthermore, let us set
\begin{displaymath}
  \zeta_{a}=e^{i\,\alpha\,\eta_{a}},\,\,\zeta_{b}
  = e^{i\,\beta\,\eta_{b}},\quad\textrm{where~}\eta_{a},
  \eta_{b}\in\{0,2\,\pi,-2\,\pi\}.
\end{displaymath}
Then one has
\begin{eqnarray*}
  & & {\KK'}^\Lambda(t,x,x_{0})\\
  & & \quad=\, 
  \zeta_a\zeta_b\,\frac{1}{4\pi it}\,
  \exp\!\left(i\,\frac{|x-x_{0}|^{2}}{4t}\right)\\
  & & \quad\quad -\,\zeta_a\,\frac{\sin(\pi\alpha)}{4\pi^2i}
  \int_{0}^{\infty}\frac{\mathrm{d}t_1}{t_1}
  \int_{0}^{\infty}\frac{\mathrm{d}t_0}{t_0}\,
  \delta(t_1+t_0-t)\\
  && \quad\qquad\qquad\qquad\qquad
  \times\,\exp\!\left(i\left(\frac{r_a^{\,2}}{4t_1}
      +\frac{r_{0a}^{\,2}}{4t_0}\right)\right)
  \frac{\exp[-\alpha(s_a-i(\theta_a-\theta_{0a}-\eta_a)]}
  {1+\exp(-s_a+i\theta_a-i\theta_{0a})}\\
  & & \quad\quad -\,\zeta_b\,\frac{\sin(\pi\beta)}{4\pi^2i}
  \int_{0}^{\infty}\frac{\mathrm{d}t_1}{t_1}
  \int_{0}^{\infty}\frac{\mathrm{d}t_0}{t_0}\,
  \delta(t_1+t_0-t)\\
  && \quad\qquad\qquad\qquad\qquad
  \times\,\exp\!\left(i\left(\frac{r_b^{\,2}}{4t_1}
      +\frac{r_{0b}^{\,2}}{4t_0}\right)\right)
  \frac{\exp[-\beta(s_b-i(\theta_b-\theta_{0b}-\eta_b)]}
  {1+\exp(-s_b+i\theta_b-i\theta_{0b})}\\
  &  & \quad\quad +\,\frac{1}{4\pi i}
  \sum_{\overline{\gamma},n\geq2}(-1)^{n}
  \int_{0}^{\infty}\frac{\mathrm{d}t_{n}}{t_{n}}
  \ldots\int_{0}^{\infty}\frac{\mathrm{d}t_{0}}{t_{0}}\,
  \delta(t_{n}+\ldots+t_{0}-t)\\
  &  & \quad\qquad\qquad\qquad\qquad
  \times\,\exp\!\left(\frac{i}{4}
    \left(\frac{r^{2}}{t_{n}}+\frac{\varrho^2}{t_{n-1}}+\ldots
      +\frac{\varrho^2}{t_1}+\frac{r_{0}^{2}}{t_{0}}\right)\right)
  S_{\overline{\gamma}}(s,\theta,\theta_{0}),
\end{eqnarray*}
where
\begin{eqnarray*}
  S_{\overline{\gamma}}(s,\theta,\theta_{0})
  & = & \frac{\sin(\pi\sigma_{n})}{\pi}
  \frac{\exp[-\sigma_{n}(s_{n}-i\theta)]}
  {1+\exp(-s_{n}+i\theta)}
  \frac{\sin\pi\sigma_{n-1}}{\pi}
  \frac{\exp(-\sigma_{n-1}s_{n-1})}{1+\exp(-s_{n})}\\
  &  & \times\ldots\times\,
  \frac{\sin(\pi\sigma_{1})}{\pi}
  \frac{\exp[-\sigma_{1}(s_{1}-i\theta_{0})]}
  {1+\exp(-s_{1}+i\theta_{0})},
\end{eqnarray*}
and
\begin{displaymath}
  s_{a} = \log\!\left(\frac{t_{1}r_{0a}}{t_{0}r_{a}}\right),
  \textrm{~}
  s_{b} = \log\!\left(\frac{t_{1}r_{0b}}{t_{0}r_{b}}\right),
  \textrm{~}
  s_{j} = \log\!\left(\frac{t_{j}r_{j-1}}{t_{j-1}r_{j}}\right)
  \textrm{~for~}1\leq j\leq n.
\end{displaymath}
In addition, $(r,\theta )$ are the polar coordinates of the point $x$
with respect to the center $c_{n}$, $(r_{0},\theta _{0})$ are the
polar coordinates of the point $x_{0}$ with respect to the center
$c_{1}$. The sum $\sum_{\overline{\gamma},\,n\geq 2}$ runs over all
finite alternating sequences of length at least two,
$\overline{\gamma}=(c_1,\ldots,c_{n})$, such that for all $j$,
$c_{j}\in \{a,b\}$, $c_{j}\neq{}c_{j+1}$, and $\sigma_{j}=\alpha$
(resp.  $\beta $) depending on whether $c_{j}=a$ (resp. $b$).

\section*{Acknowledgments}

One of the authors (P.\v{S}.) wishes to acknowledge gratefully partial
support from grant No. 201/05/0857 of the Grant Agency of the Czech
Republic.



\begin{thebibliography}{14}

\bibitem{schulman1} Schulman~L~S 1971 Approximate topologies
  \textit{J.~Math. Phys.} \textbf{12} 304-308

\bibitem{schulman2} Schulman~L~S 1981 \emph{Techniques and Applications
    of Path Integration} (New York: Wiley)

\bibitem{atiyah} Atiyah~M~F 1976 Elliptic operators, discrete groups
  and von Neumann algebras \textit{Ast{\'e}risque} \textbf{32-33}
  43-72

\bibitem{sunada} Sunada~T 1988 Fundamental groups and Laplacians
  \emph{Geometry and analysis on manifolds}, Lect. Notes Math.
  \textbf{1339} (Berlin: Springer) pp. 248-277

\bibitem{aschetal} Asch~J, Over~H and Seiler~R 1994 Magnetic Bloch
  analysis and Bochner Laplacians
  \textit{J.~Geom. Phys.} \textbf{13} 275-288

\bibitem{gruber1} Gruber~M~J 2000 Bloch theory and quantization of
  magnetic systems
  \textit{J.~Geom. Phys.} \textbf{34} 137-154

\bibitem{gruber2} Gruber~M~J 2001 Noncommutative Bloch theory
  \textit{J.~Math. Phys.} \textbf{42} 2438-2465

\bibitem{kocabovastovicek} Koc\'abov\'a~P and
  \v{S}\v{t}ov\'\i\v{c}ek~P 2008 Generalized Bloch analysis and
  propagators on Riemannian manifolds with a discrete symmetry
  \textit{J.~Math. Phys.} (to appear)

\bibitem{pla89} \v{S}\v{t}ov\'\i\v{c}ek~P 1989 The Green function for
  the two-solenoid Aharonov-Bohm effect
  \textit{Phys. Lett.~A} \textbf{142} 5-10

\bibitem{mashkevichetal} Mashkevich~S, Myrheim~J and Ouvry~S 2004
  Quantum mechanics of a particle with two magnetic impurities
  \textit{Phys. Lett.~A} \textbf{330} 41-47

\bibitem{hannaythain} Hannay~J~H and Thain~A 2003 Exact scattering
  theory for any straight reflectors in two dimensions
  \textit{J.~Phys. A: Math. Gen.} \textbf{36} 4063-4080

\bibitem{giraudetal} Giraud~O, Thain~A and Hannay~J~H 2004 Shrunk loop
  theorem for the topology probabilities of closed Brownian (or
  Feynman) paths on the twice punctured plane
  \textit{J.~Phys. A: Math. Gen.} \textbf{37} 2913-2935

\bibitem{shtern} Shtern~A~I 2001 Unitary representation of a
  topological group \textit{The Online Encyclopaedia of Mathematics}
  (Berlin: Springer), Online: http://eom.springer.de/

\bibitem{thoma} Thoma~E 1964 \"Uber unit\"are Darstellungen
  abz\"albarer, diskreter Gruppen
  \textit{Math. Annalen} \textbf{153} 111-138

\bibitem{talamanca} Fig\`{a}-Talamanca~A and Picardello~M~A 1982
  Spherical functions and harmonic analysis on free groups
  \textit{J.~Func. Anal.} \textbf{47} 281-304

\bibitem{hoermander} H\"ormander~L 2003 \emph{The Analysis of Linear
    Partial Differential Operators I} (Berlin: Springer)

\bibitem{wuyang} Wu~T~T and Yang~C~N 1978 Concept of nonintegrable
  phase factors and global formulation of gauge fields
  \textit{Phys. Rev.~D} \textbf{12} 3845-3857

\end{thebibliography}
\end{document}